\documentclass[aps,prx,twocolumn,superscriptaddress]{revtex4-2}%nofootinbib
\usepackage{amssymb,graphicx,color}
\usepackage[intlimits]{amsmath}
\usepackage[english]{babel}
\usepackage[colorlinks]{hyperref}
\hypersetup{
colorlinks=blue,
linkcolor=blue,
filecolor=blue,
urlcolor=blue,
citecolor=blue
}
\pdfoutput=1
\graphicspath{{figures/}} 
\usepackage[normalem]{ulem}
\usepackage{comment}
\usepackage{ragged2e}
\usepackage{enumerate}
\usepackage[dvipsnames]{xcolor}

\begin{document}
%\title{Interpretable data-driven approach to quantum transport}
%\title{Diagnosing quantum transport and its complexity from wave function snapshots}
\title{Diagnosing quantum transport from wave function snapshots}

%\title{Probing quantum transport from wave function snapshots}
\author{Devendra Singh Bhakuni}
\affiliation{The Abdus Salam International Centre for Theoretical Physics (ICTP), Strada Costiera 11, 34151 Trieste, Italy}
\author{Roberto Verdel}
\affiliation{The Abdus Salam International Centre for Theoretical Physics (ICTP), Strada Costiera 11, 34151 Trieste, Italy}
\author{Cristiano Muzzi}
\affiliation{SISSA — International School of Advanced Studies, via Bonomea 265, 34136 Trieste, Italy}
\affiliation{INFN, Sezione di Trieste, via Valerio 2, 34127 Trieste, Italy}
\author{Riccardo Andreoni}
\affiliation{The Abdus Salam International Centre for Theoretical Physics (ICTP), Strada Costiera 11, 34151 Trieste, Italy}
\affiliation{SISSA — International School of Advanced Studies, via Bonomea 265, 34136 Trieste, Italy}
\author{Monika Aidelsburger}
\affiliation{Max-Planck-Institut f\"ur Quantenoptik, 85748 Garching, Germany}
\affiliation{Faculty of Physics, Ludwig-Maximilians-Universit\"{a}t M\"{u}nchen, Schellingstr. 4, D-80799 Munich, Germany}
\affiliation{Munich Center for Quantum Science and Technology (MCQST), Schellingstr. 4, D-80799 Munich, Germany}
\author{Marcello Dalmonte}
\affiliation{The Abdus Salam International Centre for Theoretical Physics (ICTP), Strada Costiera 11, 34151 Trieste, Italy}

\begin{abstract}
We study nonequilibrium quantum dynamics of spin chains by employing principal component analysis (PCA) on data sets of wave function snapshots and examine how information propagates within these data sets. The quantities we employ are derived from the spectrum of the sample second moment matrix, built directly from data sets. 
Our investigations on several interacting spin chains featuring distinct spin or energy transport reveal that the growth of data information spreading follows the same dynamical exponents as that of the underlying quantum transport of spin or energy. Specifically, our approach enables an easy, data-driven, and importantly interpretable diagnostic to track energy transport with a limited number of samples, which is usually challenging without any assumption on the Hamiltonian form. These observations are obtained at a modest finite size and evolution time, which aligns with experimental and numerical constraints.
Our framework directly applies to experimental quantum simulator data sets of dynamics in higher-dimensional systems, where classical simulation methods usually face significant limitations and apply equally to both near- and far-from-equilibrium quenches.  
\end{abstract}

\maketitle

\section{Introduction}

Over the last decade, the application of data science methods to physical phenomena has witnessed a significant surge across diverse fields~\cite{carleo2019machine}. Within the realm of condensed matter and statistical physics, unsupervised and non-parametric learning techniques have been utilized to diagnose physical phenomena, such as identifying phase transitions and critical behaviour~\cite{PhysRevB.94.195105, PhysRevE.95.062122, PhysRevE.96.022140, Rodriguez-Nieva2019, PhysRevLett.125.225701, PhysRevX.11.011040}, clarifying the complexity of the problem by extracting the relevant degree of freedom without relying on prior knowledge of an \emph{order parameter}. While by now well established at equilibrium, the applicability of these techniques beyond that has just started being explored~\cite{bohrdt2021analyzing,  PhysRevE.103.052140, Shen2022,verdel2024data, Tang2024, muzzi2024principal, zang2024machine}. Therefore, it becomes crucial to delve into the potential of such methods in more intricate nonequilibrium setups. 

Unlike equilibrium systems, where information is encoded into partition functions, out-of-equilibrium systems~\cite{Polkovnikov2011Colloquium, Vasseur2016nonequilibrium,eisert2015quantum} pose greater complexity and simultaneously exhibit a wealth of fundamental phenomena such as dynamical phases of matter~\cite{khemani2016phase,else2016floquet,else2020discrete,lindner2011floquet,machado2020long,rudner2013anomalous,wintersperger2020}, anomalous quantum transport, and universal dynamics~\cite{kardar_dynamic_1986,muller1988anomalous,gerling1989anomalous,gerling1990time,agarwal_anomalous_2015,bar_lev_absence_2015,cai_1_2022,cao_entanglement_2019,corwin_kardar-parisi-zhang_2016,spohn2014nonlinear,das2020nonlinear,dupont_universal_2020,fontaine_kardarparisizhang_2022,ye2022universal,fujimoto_family-vicsek_2020,castro2016emergent,bhakuni2024dynamic, wei2022quantum, Scheie2021, Jepsen2020, Keenan2023,hopjan2023scale,hopjan2023scaleinvariant,mcroberts2022anomalous,mcroberts2022long}. Most importantly, such classes of phenomena are immediately accessible to quantum simulators and computers, where out-of-equilibrium dynamics are often easier to access than equilibrium. These settings are also particularly attractive from a data-science viewpoint, as experimenters are now capable of generating snapshots of the full many-body wave function via projective measurements~\cite{Gross2021, impertro2023local, Bernien2017, wei2022quantum,joshi2022observing,  wienand2023emergence,browaeys2020many}-- which are, however, often analyzed by relying on specific local and few-body observables. This requires specific {\it a priori} knowledge of the underlying dynamics and discards an uncontrolled amount of -- potentially critical -- information.
Given that such wave function snapshots encode extensive information about the entire many-body state, several questions arise: Can we uncover essential physics \emph{solely} from such observations without presuming an order parameter? How does information propagate within this data, and how does it relate to dynamic phenomena like quantum transport? And, importantly, can we achieve this with assumption-free and yet interpretable methods?

In this study, we address these questions by presenting a technique for extracting relevant physics from snapshots of wave functions during real-time dynamics. Utilizing principal component analysis (PCA)~\cite{jolliffe2016principal} on collections of these snapshots, we study the dynamics of information propagation within the data sets and their connection with the underlying quantum transport. The simplicity of PCA, which builds on the singular value decomposition (SVD)~\cite{jolliffe2016principal}, enhances interpretability: this enables a direct link between observations and underlying physical phenomena that we elucidate analytically for spin-1/2 systems. Intuitively, the PCA highlights the most significant degrees of freedom, thereby genuinely identifying transport carriers if present on the analyzed basis.

The first model we treat as a paradigm for spin transport is the one-dimensional XXZ chain~\cite{zotos1999finite,vznidarivc2011spin,ljubotina2017spin,ljubotina2019kardar, sarang2019kinetic, bulchandani2018Bethe,deNardis2018hydrodynamics, Dupont2021Spatiotemporal,deNardis2020Superdiffusion, Dupont2020universal,deNardis2019Anomalous, Ilievski2018Superdiffusion,denardis2023nonlinear,wei2022quantum,rosenberg2024dynamics, Ilievski2021Superuniversality,gopalakrishnan2024distinct, Krajnik2022absence,krajnik2024dynamical,ljubotina2019ballistic, Bulchandani_2021,nandy2023spin}. We observe that the Shannon entropy derived from the PCA spectrum (that we denote as PCA entropy), and the information transfer defined as the difference between largest PCA eigenvalue of the left and right data sets increase with the same dynamical exponent as the underlying spin transport. Specifically, in the easy-plane case, we observe a linear growth of PCA entropy and the information transfer, whereas in the isotropic and easy-axis cases, we observe sub-linear growth with distinct dynamical exponents consistent with the transport exponent. Furthermore, we establish an analytical link between the information transfer and the polarization transfer used in experiments to probe the nature of spin transport~\cite{wei2022quantum}.

For energy transport, we study various quantum spin chains such as Ising~\cite{kim2013ballistic}, XYZ~\cite{schulz2018energy}, and PXP models~\cite{ljubotina2023superdiffusive}, where energy is the sole conserved quantity, resulting in diffusive, ballistic, and super-diffusive energy transport, respectively. Remarkably, we observe that the PCA entropy always grows with the same dynamical exponent as the energy transport, even at timescales where the conventional measure, i.e., the energy transfer, does not yet describe the correct universality class. These observations establish a link between information spreading within the datasets and energy transport dynamics. Our analysis offers an alternative, practical approach to investigate energy transport--a task typically challenging to achieve experimentally, in the absence of major assumptions on the Hamiltonian form. 

Lastly, we incorporate random spin-flip imperfections into the data sets, often stemming from measurement errors in experiments. PCA entropy remains robust against such imperfections, accurately predicting the correct dynamical spin or energy exponent even with a moderate amount of spin-flip errors. This resilience underscores the reliability and applicability of PCA in analyzing quantum transport phenomena in realistic experimental conditions.

\begin{figure}[t]
    \centering
    \includegraphics[width=1.\linewidth]{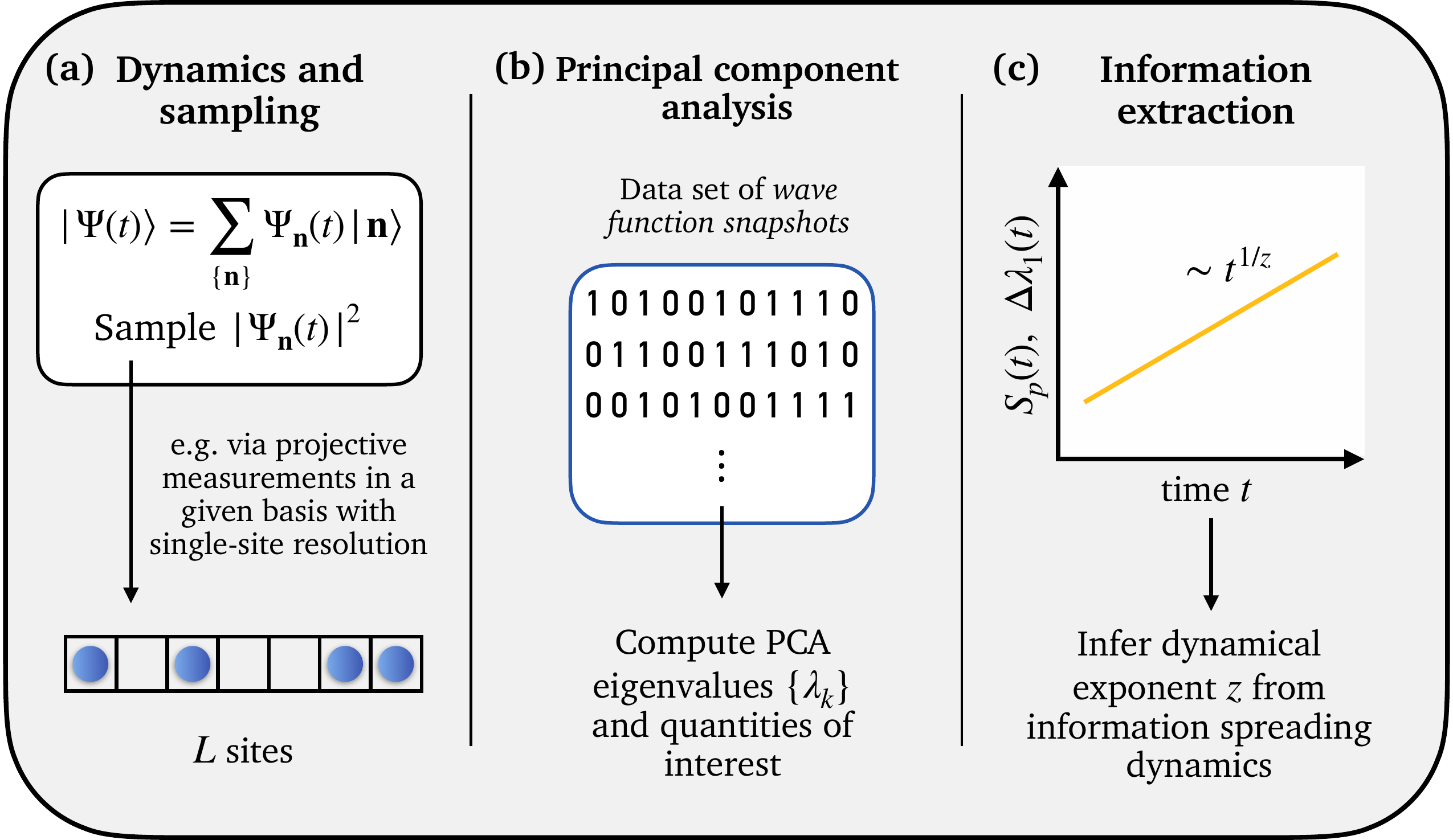}
    \caption{{Schematic of our data-driven framework for diagnosing quantum transport. (a) We consider quantum many-body dynamics with an initial imbalance in a global conserved quantity $\mathcal{Q}$. At each evolution time $t$, the many-body wave function is sampled, e.g., via projective measurements in a given basis. (b) Data sets of the sampled \emph{wave function snapshots} are then subject to a principal component analysis. (c) The dynamics of PCA spectral quantities, such as the entropy $S_p$ and the largest eigenvalue $\lambda_1$ - quantifying
    information propagation - are governed by the relevant dynamical exponent, allowing us to diagnose the nature of transport without prior knowledge of an order parameter. }}
    \label{sch}
\end{figure}

The paper is organized as follows. We provide a general workflow and define the quantities of interest in Sec.~\ref{sec:workflow}, followed by the applicability of the method for spin transport in Sec.~\ref{sec:spin} and energy transport in Sec.~\ref{sec:energy}. In Sec.~\ref{sec:errors}, we discuss the effect of imperfect measurements.  Finally, we summarize our findings in Sec.~\ref{sec:conclusions}.

\section{General workflow: data pipeline}\label{sec:workflow}
The workflow outlined in this study comprises three primary steps: (i) dynamical evolution and wave function sampling, (ii) data analysis based on PCA, and (iii) extraction of the relevant physics (Fig.~\ref{sch}). 

{\it (i) Dynamical protocol and sampling.- } The first step involves evolving a quantum many-body state under a given Hamiltonian $H$ that describes an interacting  quantum lattice model defined on $L$ sites, with some global conserved quantity $\mathcal{Q}$. In our protocol, we imprint an imbalance of the conserved quantity $\mathcal{Q}$ in the initial state, a standard approach in transport studies~\cite{ljubotina2017spin,Varma2017energy, wei2022quantum}. For simplicity, in the following we only consider spin-1/2 (qubit) systems, although our approach can be extended to systems with a larger local Hilbert space dimension~\footnote{We note that the analysis can be immediately extended in the case of other types of evolution (e.g., Floquet or open quantum systems such as boundary driven chains).}. 

The system dynamics are probed at selected evolution times via projective measurements on a given basis with single-site resolution. The outcome of such measurements are bit strings ${\bf n}=(n_1,\dots, n_L)$, with $n_j \in \{0,1\}$ for $j=1, \dots, L$, which are dubbed \emph{wave function snapshots}~\footnote{In classical simulations, wave function snapshots can be sampled, for instance, via perfect sampling~\cite{Ferris2012perfect} in matrix product state based computations.}. 
For each evolution time, the sampled wave function snapshots are organized as the row vectors of a rectangular data matrix $\mathbf{X}(t)=\{{\bf n}^{(i)}(t)\}$, where  $i=1, \dots, N_r$, labels different realizations.
Importantly, specific knowledge of the Hamiltonian $H$ is, in fact, not needed as long as we have access to the relevant wave function snapshots.  

{\it (ii) Data analysis by principal components.- } Each data matrix  $ {\bf X} $ (to simplify the notation below, we omit the explicit time dependence unless needed) is then analyzed within the mathematical framework of PCA.  
We perform an eigenvalue decomposition of an $L\times L$, symmetric matrix 
\begin{equation}
    \label{eq:cov_matrix}
    \mathbf{\Sigma}=\frac{1}{N_{r}} \mathbf{X}^{T}\mathbf{X}.
\end{equation} 
We sort the eigenvalues of this matrix in non-increasing order $\lambda_{1}\geq\dots \geq \lambda_R >0$, where $R$ is the rank of both  $\mathbf{\Sigma}$ and $\mathbf{X}$. In practice, this is conveniently done via an SVD of $\mathbf{X}$, with $\lambda_k$ determined from the corresponding singular value $s_k$ of $\mathbf{X}$, through $\lambda_k = \frac{1}{N_r}s_k^2$.  The quantities of interest are then defined from the set of eigenvalues $\{\lambda_k\}$. We note that the PCA algorithm above differs from its standard version~\cite{jolliffe2016principal} in that no centering of the data (removing the sample mean of each column from its entries) is performed. While this implies that the result of our analysis is sensitive to the choice of classical encoding of the observations $\{ {\bf n}^{(i)}\}$, as discussed later on, the results presented below are, in general, robust for sensible choices of such encoding.

Throughout our work, we primarily concentrate on the \emph{PCA entropy}~\cite{verdel2024data, 10.21468/SciPostPhysCore.6.4.086, 2023arXiv231116050V}, which is a Shannon entropy defined on the normalized spectrum of  $\mathbf{\Sigma}$, namely,
\begin{equation}
S_{\text{p}}=-\sum_{k=1}^{R}\overline{\lambda}_{k}\text{ln}(\overline{\lambda}_{k}),
\end{equation}
where $\overline{\lambda}_{k}=\lambda_{k}/\sum_{l}\lambda_{l}$. Note that the set  $\{\overline{\lambda}_{k}\}$ formally defines a probability distribution as $\overline{\lambda}_{k}$ are non-negative and sum to one. Additionally, we examine the behavior of the largest eigenvalue $\lambda_1$, which measures the \emph{variability} (i.e., statistical information) of the data along the first ``principal direction'', which is determined by the corresponding eigenvector $\vec{v_1}$. 
Both of these metrics, in their respective ways, quantify the spread of information within the wave function data sets.

{\it (iii) Information extraction.- } Finally, physically relevant information, such as dynamical exponents, is extracted from the temporal scaling of these quantities. Given that PCA identifies the most pertinent degree of freedom to characterize the data, the question arises whether transport, in terms of data, can be considered ``simple''. If so, the analysis based on PCA should yield the correct transport exponents corresponding to the conserved quantity $\mathcal{Q}$. We hypothesize that this is the case due to conserved quantities dominating the constraints in the data set, implying that both PCA entropy and the largest eigenvalue $\lambda_1$ will provide valuable insights. The remainder of the paper will scrutinize our conjecture for spin and energy transport in various quantum spin models, and provide analytical arguments in support of that.

\begin{figure*}
\includegraphics[scale=0.86]{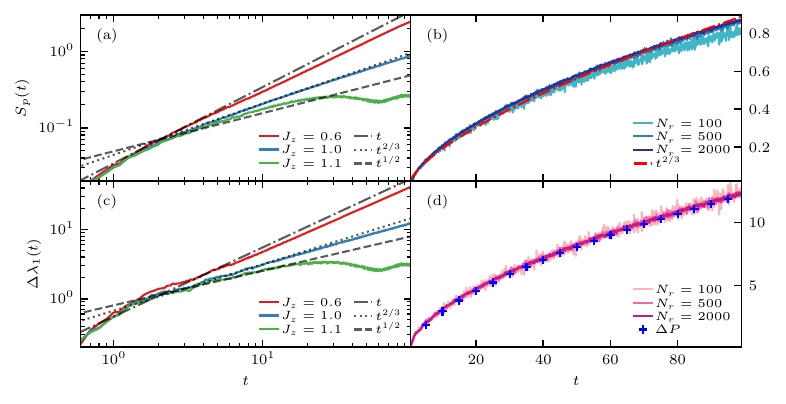}\hspace{-5pt}
\includegraphics[scale=0.86]{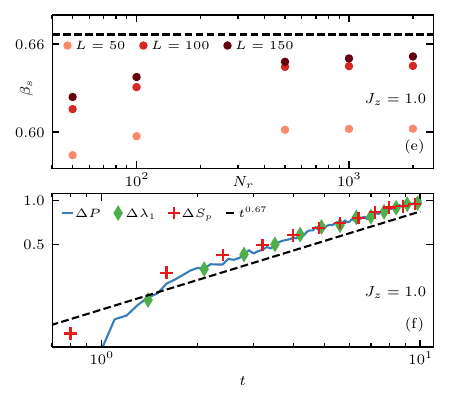}
\caption{\textit{Spin transport:} (a--d) Dynamics of the PCA entropy $S_{p}$ and information transfer $\Delta\lambda_{1}(t)$  for different parameters starting from a
domain wall for $L=150$ and $J=1$. (a,c) $S_{p}$ and $\Delta\lambda_{1}(t)$ on varying the interaction
strength $J_{z}$ at fixed $N_r=2000$. (b,d) $S_{p}$ and $\Delta\lambda_{1}(t)$ on varying the number of samples $N_{r}$ at fixed $J_{z}=1.0$. The blue markers in (d) corresponds to the polarization transfer. (e) The dynamical exponent $\beta_s$ as a function of $N_r$ extracted from $S_p(t)$ for $J_z=1.0$. (f) The dynamics of \emph{normalized} polarization transfer $\Delta P$, information transfer $\Delta\lambda_{1},(t)$ and entropy transfer $\Delta S_p(t)$ starting from an infinite-temperature state. The dashed/dotted lines provide a guide to different transport regimes.}
\label{fig1}
\end{figure*}

\section{Spin transport}\label{sec:spin}
We first demonstrate the application of the above methodology to study spin transport. We consider the one-dimensional (1D) XXZ chain of $L$ sites as a prototypical model with the Hamiltonian given by
\begin{align}
H= & \sum_{i}J(S_{i}^{x}S_{i+1}^{x}+S_{i}^{y}S_{i+1}^{y})+J_{z}\sum_{i}S_{i}^{z}S_{i+1}^{z}.
\end{align}
Here, $S_{i}^{k}, k=x,y,z$ are spin-$1/2$ operators at site $i$, $J$ is the strength of the interaction in the XY plane, which we set to $J=1$ in all our numerical simulations, and $J_z$ is the anisotropy parameter. This model conserves the total magnetization along the $z$ axis, i.e., $[H,\sum_i S_{i}^{z}]=0$, and thus allows the study of spin transport~\cite{zotos1999finite,vznidarivc2011spin,ljubotina2017spin,ljubotina2019kardar, sarang2019kinetic, bulchandani2018Bethe,deNardis2018hydrodynamics, Dupont2021Spatiotemporal,deNardis2020Superdiffusion, Dupont2020universal,deNardis2019Anomalous, Ilievski2018Superdiffusion,denardis2023nonlinear,wei2022quantum,rosenberg2024dynamics, Ilievski2021Superuniversality,gopalakrishnan2024distinct, Krajnik2022absence,krajnik2024dynamical,ljubotina2019ballistic, Bulchandani_2021,nandy2023spin}.

The behavior of spin transport and universality classes can be obtained by focusing on the infinite-temperature correlation functions~\cite{ljubotina2019kardar,ljubotina2023superdiffusive,bar_lev_absence_2015,agarwal_anomalous_2015,dupont_universal_2020} or by studying the dynamics of starting from an initial inhomogeneous state having opposite magnetization in the two halves of the chain~\cite{ljubotina2017spin, Varma2017energy}. Such an initial state can be written as  
\begin{equation}
    \label{eq:initial_state}
    \rho(0) \sim (1+\eta \sigma^z)^{\otimes L/2} \otimes (1-\eta \sigma^z)^{\otimes L/2},
\end{equation}
where the parameter $\eta$ controls the magnetization of the chain with the two halves corresponding to the magnetization $\pm \eta/2$. In the limit of $\eta=1$, the initial state reduces to a perfect domain wall state. 

The nature of the transport is then characterized by looking at the polarization transfer between the two halves and is defined as~\cite{ljubotina2019kardar,wei2022quantum} 
\begin{equation}
\label{eq:polarization_transfer}
\Delta P(t)=2\ [P^{L}(t)-P^{R}(t)].
\end{equation}
%\rv{Check factors of 2 in this equation and text below.}
Here $P^{L/R}$ is the total magnetization variation over time of the left and right half of the chain, respectively, and is given by
\begin{equation}
P^{L/R}(t)=\sum_{i\in L/R}[\langle S_{i}^{z}(t)\rangle-\langle S_{i}^{z}(0)\rangle]/2.
\end{equation}
The polarization transfer grows as a power law $\Delta P(t)\sim t^{1/z}$, with $z$ being the dynamical exponent, distinguishing the different transport regimes. For the 1D XXZ model, at the isotropic point $J_{z}=1$, the infinite-temperature transport is known to be super-diffusive, with a dynamical exponent $z=3/2$ corresponding to the Kardar-Parisi-Zhang universality class~\cite{ljubotina2017spin,ljubotina2019kardar,wei2022quantum}. On the contrary, for the easy-plane
($J_{z}<1$) and easy-axis ($J_{z}>1$) regime, the transport is ballistic and diffusive, respectively~\cite{znidaric_diffusive_2016} with the dynamical exponent being $z=1$ and $z=2$, respectively.  It is worth mentioning that for the pure domain wall $\eta=1$ and at the isotropic point $J_z=1$, the dynamics cross over to diffusion at long times~\cite{Misguich2017dynamics, Ye2020emergent}. However, in the experimentally accessible times, the dynamics remain superdiffusive~\cite{wei2022quantum}. Similarly, for $J_z\gg J$, the dynamics from a domain-wall initial state are also frozen. Here we will focus on values of $J_z \gtrsim J$, for which the polarization transfer exhibits transient dynamics. 

In the following, we focus on both the pure domain wall initial state and the partially polarized domain wall initial state. To simulate the system dynamics, we employ the time-evolving block decimation (TEBD) method~\cite{PhysRevLett.91.147902, PhysRevLett.93.040502} for a system size up to $L=150$ and with a truncation error of $10^{-8}$. We built the data matrix by a perfect sampling~\cite{stoudenmire2010minimally,Ferris2012perfect} of a matrix product state representation of the time-evolved state at different times, starting from a given initial state.

\subsection{Pure domain wall dynamics }
We first consider the case of pure domain wall initial state [$\eta=1$ in Eq.~\eqref{eq:initial_state}] and study the dynamics of the information propagation at the data set level.
\paragraph*{PCA entropy.-} We plot the dynamics of the PCA entropy $S_{p}(t)$ in Fig. \ref{fig1}(a) for various interaction strengths $J_{z}=0.5,\ 1.0,\ 1.1$, corresponding to ballistic, super-diffusive and diffusive transport for $N_{r}=2000$ samples per evolution time. As a key observation, we find that the PCA entropy grows as a power law in time $S_{p}\sim t^{1/z_{s}}$, and captures the correct dynamical exponents $z_{s}\approx z$, for all of the three cases. Dashed lines in this figure show the power law growth with the dynamical exponent $z$. 

To investigate the dependence on the number of samples, we fix the interaction strength $J_{z}=1$ and plot in Fig. \ref{fig1}(b) the dynamics of $S_{p}$ for various values of $N_{r}=100-2000$. We see that even for a small number of samples $N_{r}\sim 100$, $S_{p}$ shows a power law growth with the dynamical exponent $z_{s}$ close to $3/2$. 

A more systematic analysis of the dynamical exponent $\beta_s = 1/z_s$ extracted from the PCA entropy is plotted in Fig.~\ref{fig1}(e) as a function of $N_r$ and for a range of system sizes $L=50-150$. For larger system size and number of samples, the exponent approaches the correct dynamical exponent $\beta = 1/z$ (marked by the dashed line). Moreover, even for a small number of samples ($N_r=100$), the exponent is also quite close to the dynamical exponent $z$, suggesting that a limited number of samples is enough to obtain the information about the underlying transport. We note that even the larger volumes are still a few percent off the expected decay: such accuracy has to be expected, as we are utilizing data over timescales of order 100 $J^{-1}$, which, likely, limits our accuracy at the percent level. Additionally, the other source of this departure from the KPZ dynamical exponent could be because the superdiffusive behavior crosses over to diffusion at a late time for the domain wall initial state~\cite{Misguich2017dynamics, Ye2020emergent}. We do not observe a clear scaling with $N_r$, but we expect, at least for larger system sizes, a systematic improvement scaling as $\sqrt{N_r}$.

\paragraph*{Information transfer via largest PCA eigenvalue.-} So far, we have analyzed information from the entire wave function. We now scrutinize how information propagates between the two halves of the data set. For this purpose, we concentrate on the largest eigenvalue  $\lambda_1(t)$ of the matrix of second moments ${\bf \Sigma}$ in Eq.~\eqref{eq:cov_matrix}. 
Motivated by the experimentally relevant observable $\Delta P(t)$, we define the ``\emph{information transfer''} as 
\begin{equation}
\label{eq:information_transfer}
\Delta\lambda_{1}(t)=2[\Lambda^{L}_1(t)-\Lambda^{R}_1(t)]
\end{equation}
%\rv{Check factors of 2 to make them consistent with the definition of $\Delta P$.}
where $\Lambda^{L/R}_1(t) := - [\lambda_1^{L/R}(t) -\lambda_1^{L/R}(0) ]/2$.
Here, we first divide the data set into left and right data subsets that are
comprised of a\emph{ }$N_{r}\times L/2$ matrix each, and compute  $\lambda_{1}^{L/R}(t)$  of the left/right data subset separately. We {\it purposely} focus on single eigenvalues to test the conjecture: is spin transport a simple phenomenon (that is, something that can be accurately captured by a minimal number of variables in the context of wave function snapshots), or is it not?

In Fig.~\ref{fig1}(c), we plot the dynamics of $\Delta\lambda_{1}(t)$ as a function of time for $L=150$ and various values of the interaction strength $J_z$. Similar to the PCA entropy, the information transfer between the two data subsets happens with the same growth exponent $z$. The dependence of the information transfer on the number of samples is provided in Fig.~\ref{fig1}(d) for $J_z=1.0$. This plot also shows the dynamics of the polarization transfer $\Delta P(t)$ computed from our TEBD simulations (shown by blue cross markers).  Interestingly, we find that the information transfer $\Delta\lambda_{1}(t)$ not only provides the correct dynamical exponents for a small number of samples but also approximates the polarization transfer $\Delta P(t)$ quite well.

\subsection{Relationship between information transfer and transport}\label{subsec:XXZ_infotransfer}

The agreement between the information transfer and the polarization transfer can be understood using a fundamental theorem on SVD, which states that for any matrix $\mathbf{A}$, the sum of its squared singular values equals the square of its Frobenius norm, i.e., $\sum s_k^2(\mathbf{A}) = ||\mathbf{A}||_F^2 = \sum_{i,j} \mathbf{A}_{i,j}^2$~\cite{Blum_Hopcroft_Kannan_2020}. Since the entries of the data matrices considered here are 0 or 1, the previous result implies that
\begin{equation}
\label{eq:svd_theorem}
\sum_{k=1}^R s_k^2(\mathbf{X}) = \sum_{i,j} n_j^{(i)} = N_r \Big(\frac{L}{2} - S\Big)
\end{equation}
where $S=\frac{1}{N_r} \sum_{i=1}^{N_r}\sum_{j=1}^L \frac{1}{2}s_j^{(i)}$, with $s_j \in \{-1, 1\}$, is the average total magnetization estimated from the snapshots.  Thus, applying this result to the left and right data subsets mentioned above and the fact that $\lambda_k = s_k^2/N_r$, we obtain  
\begin{equation}
\label{eq:lambda_lr}
\lambda^{L/R}_1(t) =  \big(L/2- S^{L/R}(t)\big) - \sum_{k>1}\lambda_k^{L/R}(t).
\end{equation}
In our numerical calculations we observe that $\sum_{k> 1} \lambda_k^{L}(t) \approx \sum_{k> 1} \lambda_k^{R}(t)$, for all accessed times. Hence, when considering the difference $\lambda^{L}_1(t) - \lambda^{R}_1(t)$, such terms cancel each other out and we get
\begin{equation}
\label{eq:transfers}
\Delta \lambda_1(t) \approx \Delta P(t),
\end{equation}
with the definitions in Eqs.~\eqref{eq:polarization_transfer} and \eqref{eq:information_transfer}. We have also numerically verified this relationship for the case of ballistic and diffusive transport, respectively. 

\begin{figure}[b]
	\includegraphics[scale=1.0, width=1\columnwidth]{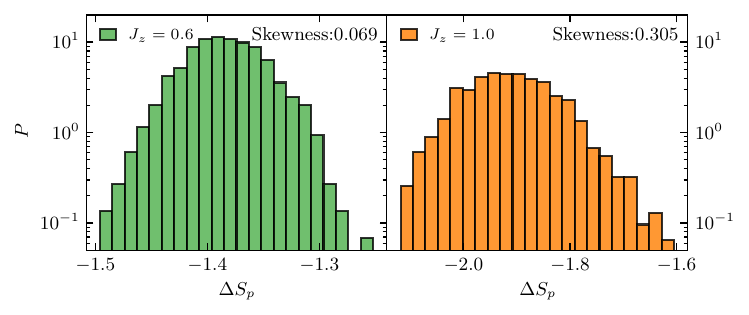}
	\caption{Probability distribution of the entropy transfer $\Delta S_p$ between the left and right data sets. Left: for ballistic transport ($J_z=0.6$), this distribution becomes symmetric with skewness 0.069. Right: for KPZ super-diffusive transport ($J_z=1.0$), the distribution becomes asymmetric with skewness 0.305.}
	\label{fig2}
\end{figure}

\subsubsection{Distribution of information and entropy transfer}\label{subsub:TW}
The above relationship also implies that the distribution of the information transfer for batches of the data set can feature the Tracy-Widom distribution, a hallmark of the KPZ universality class~\cite{wei2022quantum}. 
Interestingly, not only the information transfer but in fact, the whole \emph{entropy transfer} between the two halves of the system, which considers \emph{all} eigenvalues and is defined as
\begin{equation}
\label{eq:entropy_difference}
\Delta S_p(t) =  S_{p}^{L}(t) -  S_{p}^{R}(t),
\end{equation} 
where $S_{p}^{L/R}$ is the PCA entropy of the left and right data sets, respectively, follows the Tracy-Widom distribution. We compare the distribution of the entropy transfer for the ballistic ($J_z=0.6$) and the KPZ ($J_z=1.0$) case in Fig.~\ref{fig2}. For ballistic transport, the distribution of the entropy transfer becomes symmetric, while for $J_z=1$, we see the emergence of Tracy-Widom distribution, which confirms the KPZ super-diffusive transport at $J_z=1$.

\subsection{Infinite-temperature spin transport}
In this subsection, we consider the infinite-temperature inhomogeneous state defined in Eq.~\eqref{eq:initial_state} with $\eta=0.67$. The evolution of the mixed-density matrix is, however, challenging for larger system sizes. Instead, here, for numerical purposes, we start with ensembles of random product states given by
\begin{equation}
|\Psi(0)\rangle = \bigotimes_{i<L/2} |\psi(+\eta, \phi_i)\rangle \bigotimes_{i\geq L/2} |\psi(-\eta, \phi_i)\rangle,
\end{equation}
with $|\psi(-\eta, \phi_i)\rangle = \sqrt{(1+\eta)/2} |\uparrow\rangle + e^{i\phi_i}\sqrt{(1-\eta)/2} |\downarrow\rangle$, and $\phi_i$ are drawn randomly from the uniform distribution $\phi_i\in[0,2\pi)$. Averaging over the ensemble of these pure states is equivalent to an infinite temperature density matrix in Eq.~\eqref{eq:initial_state}~\cite{ljubotina2017spin}.

We consider $M=40$ random product states and generate $N_r$ configurations for each $\phi_i$. We build a new data matrix of size $(M N_r)\times L$ to perform the PCA. Due to the nature of the initial high-energy state, the PCA entropy $S_p(t)$ of the full data set saturates rather quickly and is unsuitable for characterizing spin transport. We, therefore, rely on the information transfer between the left and right data sets. Additionally, we also study the entropy transfer between the two halves of the chain introduced in Eq.~\eqref{eq:entropy_difference}. 

In Fig.~\ref{fig1}(f), we plot the dynamics of the normalized polarization transfer, the normalized entropy difference, and the normalized information transfer for $L=100$ and at a fixed interaction strength $J_z=1.0$. Similar to the domain-wall initial state, we see that the dynamics of these quantities grow as a power law, which is close to the KPZ dynamical exponent (shown as the dashed line) and thus highlights the applicability of our analysis for a broader class of initial states.

\subsection{Summary: Spin transport from  data sets}
In summary, by performing PCA on the data sets representing the wave function snapshots, we found that the quantities like PCA entropy and information/entropy transfer grow with the same dynamical exponent as the underlying spin transport. We show this for two different choices of initial states, namely, the perfect domain wall initial state and an inhomogenous infinite-temperature state. Additionally, we provide an analytical understanding of the observed features of the information transfer, which can be approximated to the polarization transfer.

Since the largest eigenvalue of the PCA spectrum is enough to extract the dynamical exponent of spin transport, this implies that spin transport is a fundamentally simple phenomenon at the data level and can be inferred by dimensionality reduction. 

\section{Energy transport}\label{sec:energy}
We now investigate energy transport, focusing specifically on systems with energy being the only conserved quantity and systems where energy conservation is accompanied by magnetization conservation. In general, akin to spin transport, the characteristics of energy transport can be elucidated by analyzing infinite-temperature \emph{energy-energy} correlation functions~\cite{kim2013ballistic, Varma2017energy}. However, unlike spin transport, which has been experimentally observed, probing energy transport poses significant challenges in experiments. It requires the precise form of the local energy function, which is heavily reliant on assumptions, and generically requires measurement on all possible bases, as well as inter-basis measurements. 

Following the choice of the initial state for the spin transport, we consider the initial state with an \emph{energy domain-wall}~\cite{Varma2017energy} by preparing the left half in the ground state of Hamiltonian $+H$ and the right in the ground state of the Hamiltonian $-H$, respectively~\footnote{Alternatively, a more experimentally feasible option involves a product state wherein one-half of the system is fully polarized in the $z$ direction while the other half is prepared in a N\'eel state.}. At $t=0$, the two halves of the chains are joined, and the system evolves according to the full Hamiltonian $H$. We do not make any assumption on what the specific form of the Hamiltonian is.

The nature of the energy transport can be characterized by focusing on the energy transfer between the two halves, defined as
\begin{equation}
\label{eq:energy_transport}
\Delta E(t) = E_L(t) - E_R(t),
\end{equation}
where $E_{L/R}(t)=\sum_{i\in L/R}[\langle H_{i}(t)\rangle-\langle H_{i}(0)\rangle]/2$ is the total energy of the left and the right part, respectively. Starting from the aforementioned initial state, the energy transfer grows as a power-law $\Delta E\sim t^{1/z_\epsilon}$ where the dynamical exponent $z_\epsilon$ provides information about the nature of the transport. A dynamical exponent of $z_\epsilon = 1$ signifies ballistic transport, whereas $z_\epsilon = 2$ characterizes diffusion. When $z_\epsilon < 2$, energy propagation exhibits super-diffusive behavior, whereas $z_\epsilon > 2$ indicates sub-diffusive energy transport.

To perform PCA, similar to the previous scenario, we construct the data matrix by collecting the snapshots on the $\sigma^z$ basis at different times. Notably, in the examined model, measurements in the $\sigma^z$ basis suffice to discern the hallmark of the underlying energy transport. This is one of the key observations presented in this work. A detailed exploration of alternative basis choices will be presented elsewhere.

\begin{figure*}[t]
\includegraphics[scale=1.54]{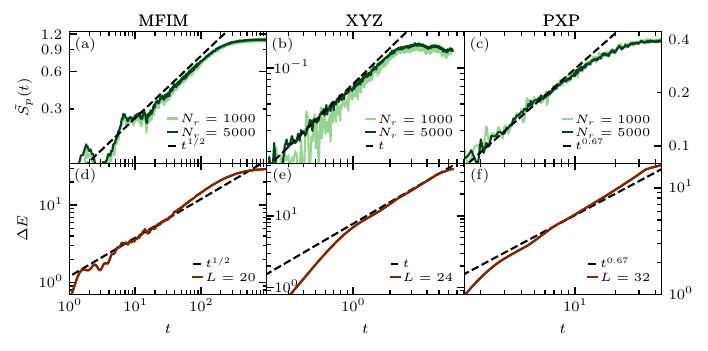}
\caption{\textit{Energy transport}: Dynamics of PCA entropy $S_p(t)$ and the energy transport $\Delta E$ starting from an initial \emph{energy domain wall} state. (a,d) For the MFIM, (b,e) for the XYZ spin chain, and (c,f) for the PXP model. The dashed lines provide a guide to the different transport regimes.}
\label{fig3}
\end{figure*}

In the following, we consider different quantum many-body systems featuring distinct energy transport.

\subsection{Energy conserving models}

We first consider systems where energy is the only conserved quantity and focus on models that exhibit different types of energy transport.

\subsubsection{Ising model} 
We consider a non-integrable one-dimensional mixed-field Ising model (MFIM) with the Hamiltonian 
\begin{align}
H= & g\sum_{i=1}^{L}\sigma_{i}^{x} + \sum_{i=2}^{L-1} h\sigma_{i}^{z} + (h-J)( \sigma_{1}^{z} + \sigma_{L}^{z})+J\sum_{i=1}^{L-1}\sigma_{i}^{z}\sigma_{i+1}^{z}.
\end{align}
Here, $g$ and $h$ correspond to the strength of the transverse and the longitudinal field, respectively, and $J$ is the strength of interaction along $z$ direction. For our numerical simulations, we fix the parameters: $J=1$, $g=(\sqrt{5}+5)/8$, and $h=(\sqrt{5}+1)/4$. In this non-integrable model, it is established that the entanglement entropy grows ballistically in time, while the energy spreading is diffusive ($\Delta E \sim t^{1/2}$)~\cite{kim2013ballistic}.

\subsubsection{XYZ model} In addition, we investigate a system wherein energy transport follows a non-diffusive behavior. To this end, we concentrate on an integrable one-dimensional XYZ spin chain, where energy transport is recognized to be ballistic~\cite{schulz2018energy}. The model Hamiltonian can be written as
\begin{align}
H= & \sum_{i}J_x\sigma_{i}^{x}\sigma_{i+1}^{x}+J_y\sigma_{i}^{y}\sigma_{i+1}^{y}+J_{z}\sigma_{i}^{z}\sigma_{i+1}^{z},
\end{align}
where $J_x=J+\delta$, $J_y=J-\delta$, and $J_z$ are the interactions along three directions and $\delta$ is the anisotropy parameter. We set $J_z, J=1$,  and $\delta=0.1$ for the numerical purposes.

\subsubsection{Kinetically constraint PXP model} Finally, we consider a model where energy transport is neither ballistic nor diffusive, but the infinite temperature energy transport is known to be super-diffusive. We consider a kinetically-constrained PXP model~\cite{Fendley2004competing,ljubotina2023superdiffusive} with the Hamiltonian given by 
\begin{align}
H= & \Omega \sum_{i}P_{i-1} X_i P_{i+1}.
\end{align}
Here, $X_i$ is the Pauli $X$ matrix, and $\Omega$ is the Rabi frequency, which we set to $\Omega=1$. The operators $P_i = \frac{1-n_i}{2}$ are the local projectors on $|$$\downarrow\rangle$ state and discard the possibility of two up-spins being next to each other.

\subsubsection{Numerical results}

For the models discussed above, we perform the PCA on the data sets generated using exact diagonalization. Due to the choice of initial state (constructed from the ground state of subsystem Hamiltonians $\pm H$), we always have a non-zero initial value of the PCA entropy. Hence, in order to filter out this entropic contribution, we focus on the dynamics of PCA entropy difference: $\tilde{S}_p(t) = S_p(t)-S_p(0)$.

We show the dynamics of $\tilde{S}_p(t)$ for various numbers of samples in Fig.~\ref{fig3} alongside the energy transport $\Delta E(t)$ [Eq.~\eqref{eq:energy_transport}]. For both the MFIM and XYZ chain, we verify that the energy transfer grows as a power law with the dynamical exponent being $z_\epsilon=2$ and $z_\epsilon=1$ [Fig.~\ref{fig3}(d,e)] corresponding to the diffusive and ballistic transport, respectively. Interestingly, similar to the energy transport, the PCA entropy difference $\tilde{S}_p(t)$ also grows as a power law $\sim t^{1/z_s}$ [Fig.~\ref{fig3}(a,b)], capturing the correct dynamical exponent $z_s\approx z_\epsilon$. While for small values of $N_r=1000$, we observe large fluctuations, increasing the number of samples (yet with $N_r\ll 2^L$) reduces the fluctuations significantly. Furthermore, we observe that the saturation of the PCA entropy and energy transfer happens at different timescales. The information spreading on the data set with snapshots only in the $\sigma^z$ basis happens faster than the energy transfer. Such difference is likely due to finite volume effects acting differently on these quantities and not to the reduced statistics of PCA entropies (changing $N_r$ has little effect on the saturation point).

Similarly, for the PXP model, we illustrate the dynamics of the PCA entropy difference and the energy transfer $\Delta E$ in Fig.~\ref{fig3} (c,f). The energy transfer grows as a power law with dynamical exponent $1/z_\epsilon \approx 2/3$ (The discrepancy from the power law of $2/3$ is due to finite-size effect, see e.g., Ref.~\cite{ljubotina2023superdiffusive} which employ MPS simulations for larger system sizes). Similar to the energy transfer, the PCA entropy difference also grows with the same dynamical exponent. Thus, an analysis based on PCA can aid in determining the accurate dynamical exponent of energy transport and, in fact, can result as far superior to energy transfer: timescales of order $10$, sample batches of order of a few thousands, and sizes of order $30$ sites are sufficient to correctly determine a dynamical scaling exponent. All of these are well attainable in Rydberg quantum simulators~\cite{browaeys2020many}.

\subsection{Energy and spin conservation}

\begin{figure}[t]
\includegraphics[scale=1.0]{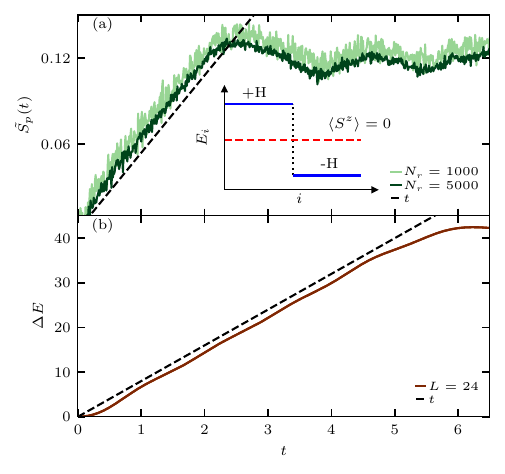}
\caption{\textit{Energy transport:} Dynamics of PCA entropy $S_p(t)$ and the energy transport $\Delta E$  in the XXZ model, starting from an initial \emph{energy domain wall} state but with a conserved homogeneous magnetization profile, as schematically shown in the inset in (a).}
\label{fig4}
\end{figure}

So far, in this section, our focus has been on systems where energy is the sole conserved quantity. Now, we aim to assess the applicability of our analysis in predicting the nature of transport in systems that conserve both energy and magnetization. To this end, we turn our attention to the 1D XYZ chain with $\delta=0$ and a fixed magnetization $\langle S^z \rangle = 0$. While spin transport exhibits super-diffusive behavior, energy transport is known to be ballistic in such systems. We examine the initial state featuring an energy domain wall but with a homogeneous magnetization profile [see inset in Fig.~\ref{fig4}(a)].

We plot the dynamics of the PCA entropy difference $\tilde{S}_p(t)$ for different numbers of samples $N_r$, and the energy transfer $\Delta E$ in Fig.~\ref{fig4}(a,b), respectively. We observe that the growth of both quantities follows a linear trend over time, indicating a ballistic nature of energy transport. Interestingly, the saturation time for information propagation on the data sets is relatively short compared to the energy transport. Thus, our analysis implies that by introducing a suitable kink in the density profile of the conserved quantity $\mathcal{Q}$ in the initial state, the analysis carried out here can effectively characterize the underlying transport of that conserved quantity $\mathcal{Q}$. 

\begin{figure}[b]
\includegraphics[scale=1.0]{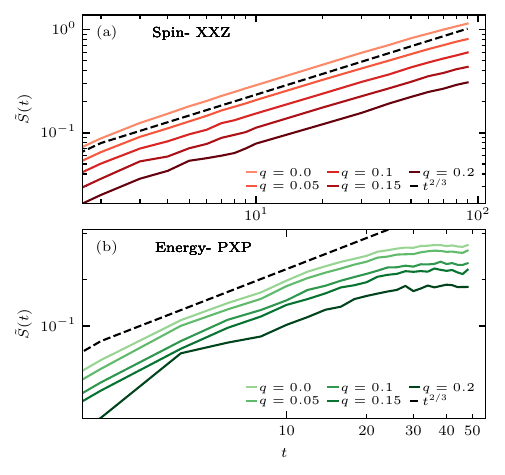}
\caption{Finite measurement errors: Dynamics of PCA entropy difference $\tilde{S}_p(t)$ for imperfect measurement with spin-flip fraction $q$ (a) for spin transport in the XXZ model and (b) energy transport in the PXP model. The dashed black line provides a guide to the power law $t^{2/3}$. }
\label{fig5}
\end{figure}

\subsection{Summary: Energy transport from data sets}

To summarize, in this section, we demonstrate the utility of PCA for characterizing energy transport in various setups, conserving only total energy or both total energy and magnetization, each with a distinct nature of energy transport. Our analysis suggests that the information about the underlying energy transport can be extracted from PCA entropy, which grows with the same dynamical exponent. In analogy to spin transport, similar results can be obtained using the largest eigenvalue, albeit with larger errors.

It is noteworthy to highlight that while spin transport has been investigated in experiments, observing energy transport presents a significant challenge. However, our findings indicate that energy transport can be deduced from wave function snapshots on the $z-$basis, a data type readily accessible in experiments that offers a promising avenue for studying energy transport in experiments. And, importantly, short times and small volumes already indicate the correct dynamical exponent, in some cases with superior accuracy with respect to the (experimentally hard to access) energy transport measures.

\section{Effect of measurement errors and encoding}\label{sec:errors}

\subsection{Resilience against spin-flip measurement errors}
In this section, we consider the effect of imperfect measurements, which typically arise in realistic experimental settings and correspond to random spin slips. In the following, we show the robustness of our method in identifying the correct transport exponent, even with these imperfect measurements.

To this end, we introduce finite spin flips on our data set at each time instance. On the data level, this amounts to changing the bit value (``0" to ``1" and vice-versa) on a fraction of sites $q$ that are chosen randomly for each snapshot. Such a flipping process introduces an extra contribution to the PCA entropy in the initial state. Therefore, we focus on the PCA entropy difference $\tilde{S}_p(t)$ introduced previously  [Eq.~\eqref{eq:entropy_difference}]. To illustrate this, we specifically focus on the spin transport in the XXZ model and the energy transport in the PXP model. We have checked that similar results hold for the energy transport in the MFIM and XYZ models. 

Fig.~\ref{fig5} depicts the dynamics of the PCA entropy difference $\tilde{S}_p(t)$ for the spin transport in the XXZ model [Fig.~\ref{fig5} (a)] and for the energy transport in the PXP model [Fig.~\ref{fig5} (b)] for various flipping fraction $q$ and with a fixed number of samples $N_r=5000$. The dynamical exponent corresponding to the spin transport in the XXZ model and the energy transport in the PXP model predict superdiffusion with $z,z_\epsilon\approx 3/2$. For both cases, $\tilde{S}_p(t)$ continues to grow as a power law even for a finite flipping fraction $q$. Notably, despite random flips for the selected probabilities, the dynamical exponents remain very close to $z, z_\epsilon \approx 3/2$. A guide to the power law $t^{2/3}$ is provided in Fig.~\ref{fig5} by a black dashed line, affirming the robustness of the PCA method in characterizing the dynamical exponent despite very considerable imperfections.

While we do not know the origin of the unexpected resilience of the protocol to measurement errors, it is worth emphasizing that, at the information theoretical level, principal component spectra are known to undergo ``learning'' transitions in specific random matrix models~\cite{doi:10.1073/pnas.2302028120}. It would be interesting to understand whether those transitions are related to the present observations and if they can provide additional knobs to control errors in case of other experimental imperfections.

\subsection{Relevance of the encoding}
%{\it Relevance of the encoding. -} 
The results mentioned above were obtained using an uncentred version of PCA. As noted before, this makes our analysis sensitive to the choice for representing the classical data ${\bf n} =(n_1, \dots, n_L)$. Even when  considering only an encoding with $n_j \in \{0, 1\}$, different numerical results are obtained upon ``reversing'' our convention, i.e., $n_j \to \tilde{n}_j = (n_j + 1) \,\ \mathrm{mod} \,\ 2$. We found, however, that unless the data matrix is highly ``homogeneous'' [i.e., containing an overwhelmingly larger number of 0's (1's) than 1's (0's)], the dynamical exponents governing the growth of information spreading in both cases are consistent with each other. 
As an empirical observation, we found better results (smaller fluctuations in the information-spreading signal), in the representation for which the data points do not concentrate very close to the origin.
We note that with standard PCA, i.e., centering the data before performing an SVD on the data matrix ${\bf X}$, we were not able to extract information about the dynamical exponents associated with quantum transport, at least in the setting considered in this work. This becomes particularly clear when considering the information transfer in the XXZ model, whose relation to the polarization transfer noted in Sec.~\ref{subsec:XXZ_infotransfer}  breaks down upon centering the data. 
We do not disregard, however, the potential usefulness of PCA with mean-centering in different scenarios where statistics around the mean play a central role, as in the case of fluctuating hydrodynamics~\cite{wienand2023emergence}. Here, on the other hand, it appears that the centroid of the data cloud itself reveals important physical information, and thus, uncentred PCA does provide informative insights.

\section{Discussion and outlook}\label{sec:conclusions}
Utilizing PCA on data sets of snapshots (in the $z$-basis) of the full many-body wave function, we demonstrate that fundamental quantities derived from the singular values of the data matrix provide knowledge about the information spreading on the data set and its connection with the underlying nature of quantum transport of conserved quantities. This suggests that transport can be viewed as an emergent and relatively simple phenomenon from the perspective of information theory, accessible via linear dimensional reduction schemes. At the conceptual level, we conclude that data information transfer is dominated by conserved quantities, a highly non-trivial fact that, for the spin-1/2 case, can be analytically connected to the specific structure of principal component analysis. Our findings are supported by numerical investigations across various interacting quantum spin chains exhibiting distinct energy or spin transport. 

In terms of application to experiments, our method capitalizes on the capabilities of quantum simulators, leveraging wave function snapshots obtained through in-situ and flexible basis imaging~\cite{Gross2021, impertro2023local, Bernien2017, wei2022quantum,joshi2022observing,  wienand2023emergence,browaeys2020many}. Importantly, our approach directly applies to experimental wave function snapshots of higher-dimensional quantum systems, where conventional numerical methods face limitations due to entanglement growth and exponential scaling of the Hilbert-space dimension. For the case of energy transport, we identify clear cases where the latter can be experimentally probed at a quantitative level, which is otherwise challenging if no input on the exact Hamiltonian functional is {\it a priori} given. Remarkably, for energy transport, we demonstrate that small volumes and reduced statistics are sufficient for accurately determining the dynamical exponents, suggesting that dimensional reduction is particularly effective for such dynamics - that is, energy transport is simple at the data structure level. This may be particularly relevant to Rydberg atom quantum simulators~\cite{browaeys2020many}, where our method predicts the correct universal properties already at small sizes, short evolution times, and at the price of modest statistics, in regimes where even energy transport measures are failing to get the correct physics.

Our work raises several intriguing questions for future exploration. For instance, it prompts investigation into whether these methods or their generalizations can help infer higher-order correlation functions~\cite{Rispoli2019}, particle number fluctuations, and surface roughness~\cite{fujimoto_dynamical_2021,fujimoto_family-vicsek_2020,bhakuni2024dynamic,wienand2023emergence,fujimoto2024exact,aditya2024family} to comprehend fluctuating hydrodynamics. Additionally, it remains to be seen whether these methods exhibit robustness against dephasing and particle losses: in principle, PCA can be complemented with specific noise-filtering methods~\cite{barbier2023fundamental}, a particularly attractive tool to mitigate the effect of dissipation effects. Another interesting perspective is to apply the methods discussed here to dynamical phenomena beyond transport, including quantum thermalization and ergodicity breaking, particularly to probe the concept of deep thermalization and complete Hilbert-space ergodicity~\cite{Choi2023,cotler2023emergent,pilatowsky2023complete, mark2024maximum,shaw2024universal}, which have strong information theoretic grounds. We leave these questions to future investigations. 

\section*{Acknowledgments} 
We thank M. Heyl, Z. Lenar\v{c}i\v{c}, M. Ljubotina, A. Scardicchio, M. Serbyn, C. Vanoni and L. Vidmar for the discussions. We also thank R. S. Cortes, A. Gambassi, A. Jelic, R. K. Panda and V. Vitale for collaboration on related projects, and the group of J. Barbier for discussions on principal component analysis theory.
M.\,D. was partly supported by the QUANTERA DYNAMITE PCI2022-132919, by the EU-Flagship programme Pasquans2, by the PNRR MUR project PE0000023-NQSTI,  and by the PRIN programme (project CoQuS). M.\,A. received funding from the
Deutsche Forschungsgemeinschaft (DFG, German Research Foundation) via Research Unit FOR5522 under
project number 499180199 and under Germany’s
Excellence Strategy – EXC-2111 – 390814868. This publication has further received fund-
ing under Horizon Europe programme HORIZON-CL4-
2022-QUANTUM-02-SGA via the project 101113690
(PASQuanS2.1). The simulations
were performed using the Julia version of ITensor~\cite{Matthew2022itensor,Matthew2022codebase} and QuSpin~\cite{philip2017quspin,philipe2019quspin}
libraries, respectively.

\bibliography{ref}

\end{document}